\documentclass[twocolumn]{jpsj3}
\usepackage{graphicx,color}

\def\diff{\mathrm d}

\title{
Electronic Properties and Persistent Spin Currents of Nanospring\\ under Static Magnetic Field
}
\author{Taichi Kosugi}

\inst{
Nanosystem Research Institute (NRI) ``RICS'', AIST, 1-1-1 Umezono, Tsukuba 305-8568, Japan \\
}

\abst{
Relativistic electronic properties of a nanospring under a static magnetic field are theoretically investigated in the present study.
The wave equation accounting for the spin-orbit interaction is derived for the nanospring
as a special case of the Pauli equation for a spin-$1/2$ particle confined to a curved surface under an electromagnetic field.
We define the helical momentum operator and show that it commutes with the Hamiltonian owing to the helical geometry of the nanospring.
The energy eigenstates are hence also the eigenstates of the helical momentum.
We solve the equation numerically to obtain the surface wave functions and the energy spectra.
The electronic properties are systematically examined by varying the parameters that characterize the system.
It is demonstrated that either the nonzero spin-orbit interaction or the nonzero external magnetic field suffices for the occurrence of the persistent spin current on the nanospring.
Two different mechanisms are shown to generate the persistent spin current.
One employs the spin-orbit interaction coming from the local inversion asymmetry on the surface,
while the other employs the curvature coupling with the external magnetic field.
}

\kword{spin current, spin-orbit interaction, nanospring}

\begin{document}
\maketitle

\section{Introduction}

Helices have been fascinating scientists for centuries not only in physics but also in other diverse fields
since helical structures are widely seen in nature.
One of the most famous examples is DNA, which has a double-helix structure formed by hydrogen bonds.
Another example is proteins, many of which contain helical substructures.
The helical structures of DNA and proteins are believed to play important fundamental roles  in modern biology.

On the other hand, the recent development of nanotechnology facilitates 
the fabrication of materials in various shapes and sizes, even including inorganic materials in helical forms.\cite{bib:2099}
They have various constituent compounds such as ZnO\cite{bib:2087,bib:2088,bib:2095,bib:2093}, SiO$_2$\cite{bib:2085,bib:2096,bib:2091}, Pd\cite{bib:2098}, SiC\cite{bib:2086}, PbSe\cite{bib:2555}, and InGaN\cite{bib:2092}.
The length scales of such inorganic helices range from nanometers to micrometers.
Since the helical form of nanomaterial allows it to be stretched or compressed without plastic elasticity if the force is not too strong,
they are of interest for mechanical applications.\cite{bib:2556}
The much more important and interesting properties for both theoretical and experimental studies are the electronic properties,\cite{bib:2557,bib:2558}
in which the helical and curved geometry should influence the behavior of electrons.
Nanostructures with curved geometry in other than helical forms\cite{bib:1685,bib:1680_3,bib:1680_4,bib:1680_9,bib:1680_13,bib:1680_15} have also been fabricated.

Curved geometry is known to induce a geometric potential\cite{bib:1649}, which affects the dynamics of an electron moving on the curved surface,
even when an electrostatic potential is absent.
The curvature effects of surfaces in a helical geometry on the nonrelativistic electronic properties
without the spin degree of freedom have been theoretically studied.\cite{bib:1652,bib:2254}
There are many studies on  examining the electronic properties of two-dimensional systems in other nontrivial geometries,
which include a plane with a bump\cite{bib:1651},
a cylinder in a transverse magnetic field\cite{bib:1646},
a catenoid\cite{bib:1954},
and a rolled-up nanotube.\cite{bib:1642}
Grigor'kin and Dunaevskii examined the electronic and optical properties of a cylinder with a helical potential in their works.\cite{bib:1844,bib:2062,bib:1845}
For one-dimensional systems,
the electronic and optical properties of a helix have been studied.\cite{bib:1696,bib:1701,bib:1702,bib:1697,bib:1698}
The torsion-induced persistent charge currents on twisted quantum wires\cite{bib:1689,bib:1647,bib:1899,bib:1896} and knotted tori\cite{bib:2549} have also been studied.
Despite the theoretical studies of the various kinds of nontrivial geometries reviewed above,
no systematic study of the curvature effects of helical geometry on the relativistic electronic properties with spin degree of freedom has been reported.

The Schr\"odinger equations for a quantum mechanical particle on a curved surface have been used
as tools for theoretical investigations on low-dimensional systems of nontrivial geometry.
One of the most reliable methods on which their formulations are based is the thin-layer method, proposed by da Costa~\cite{bib:1649}.
This method regards the curved surface as a two-dimensional system embedded in a flat three-dimensional space.
Ferrari and Cuoghi\cite{bib:1645} adopted this approach and rigorously demonstrated, by choosing a proper gauge,
that the separation of the on-surface and transverse dynamics under an electromagnetic field is possible without approximation.
Kosugi\cite{bib:1893} extended recently their Schr\"odinger equation to the Pauli equation,
which can describe a charged spin-$1/2$ particle with a nonzero mass confined to a curved surface under an electromagnetic field.

In the present study,
the relativistic electronic properties of a nanospring\cite{bib:2090} under a static magnetic field are examined.
The wave equation accounting for the curvature effects and the spin-orbit interaction (SOI) is derived for the nanospring
as a special case of the Pauli equation for curved surfaces.\cite{bib:1893}
The equation is solved numerically and the interesting phenomena occurring on the spring are analyzed in detail.

This paper is organized as follows.
In \S 2, the description of the geometry of a nanospring as a curved surface is provided by using the curvilinear coordinates.
In \S 3, the Pauli equation for the nanospring to be satisfied by the two-component wave functions is derived.
In \S 4, the electronic properties of the nanospring are systematically examined by varying the parameters that characterize the system.
In \S 5, the conclusions of the present study are provided.

\section{Geometry of Nanospring}

\subsection{Curvilinear coordinate system on nanospring}

\begin{figure}[htbp]
\begin{center}
\includegraphics[keepaspectratio,width=6cm]{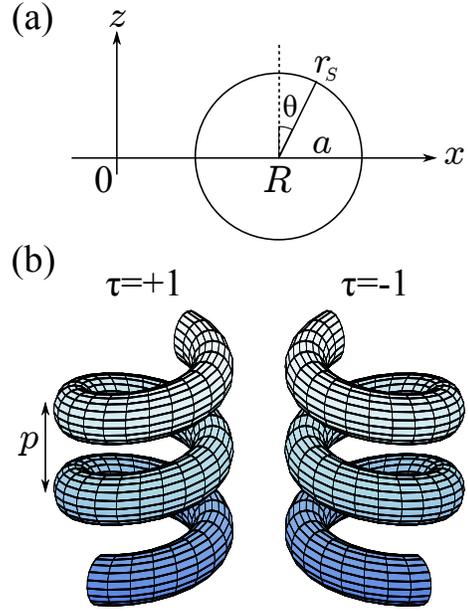}
\end{center}
\caption{
(Color online)
(a) Circle of radius $a$ on the $xz$ plane for generating a nanospring, viewed along the $y$ axis.
Its center is located at $(R, 0 ,0)$.
By rotating the circle around the $z$ axis while translating it in the $z$ direction,
the nanospring with pitch $p$ is obtained.
(b) Two examples of nanosprings with chirality $\tau = +1$ on the left and $\tau = -1$ on the right.
Solid curves on their surfaces represent constant-$\theta$ and -$\phi$ contours.
}
\label{Fig_geometry}
\end{figure}

In this subsection, the expression of the curved surface representing a nanospring is provided.
Let us consider a circle of radius $a$ on the $xz$ plane whose center is at $(R, 0, 0)$.
By rotating the circle around the $z$ axis while translating it in the $z$ direction,
we obtain a curved surface swept by the circle (see Fig. \ref{Fig_geometry}). 
Hereafter, we call it the nanospring,
which is the two-dimensional system to be investigated quantum mechanically in the present study.
An arbitrary point on the nanospring is represented by the two coordinates $\theta$ and $\phi$ as
\begin{gather}
	\boldsymbol{r}_S(\theta, \phi) = 
		\begin{pmatrix}
			(R + a \sin \theta) \cos \phi \\
			(R + a \sin \theta) \sin \phi  \\
			a \cos \theta + \tau \gamma \phi
		\end{pmatrix}
	,
	\label{spring_point}
\end{gather}
where $-\infty < \phi < \infty$ is the azimuthal angle
and $\theta$ specifies the position on the circle.
We have defined $\gamma \equiv \frac{p}{2 \pi}$, where $p > 0$ is the pitch of the nanospring.
$\tau$ specifies the chirality of the nanospring, which can take only $+1$ or $-1$.

The tangent vectors and the normal vector are given by
\begin{gather}
	\boldsymbol{e}_{\theta} \equiv \frac{\partial \boldsymbol{r}_S}{\partial \theta} = 
		a
		\begin{pmatrix}
			\cos \theta \cos \phi \\
			\cos \theta \sin \phi \\
			-\sin \theta 
		\end{pmatrix}
	, \nonumber \\
	\boldsymbol{e}_{\phi} \equiv \frac{\partial \boldsymbol{r}_S}{\partial \phi} = 
		\gamma
		\begin{pmatrix}
			-\xi \sin \phi \\
			 \xi \cos \phi \\
			\tau 
		\end{pmatrix}
	, \nonumber \\
	\boldsymbol{e}_n \equiv \frac{\boldsymbol{e}_{\theta} \times \boldsymbol{e}_{\phi}}{|\boldsymbol{e}_{\theta} \times \boldsymbol{e}_{\phi}|}
	= \frac{1}{\eta}
		\begin{pmatrix}
			\xi \sin \theta \cos \phi + \tau \cos \theta \sin \phi \\
            \xi \sin \theta \sin \phi - \tau \cos \theta \cos \phi \\
			\xi \cos \theta 
		\end{pmatrix}
	,
	\label{spring_vectors}
\end{gather}
where
$\xi \equiv \frac{R + a \sin \beta}{\gamma}$ and
$\eta \equiv \sqrt{\cos^2 \theta + \xi^2}$.
It is noted that $\boldsymbol{e}_\theta$ and $\boldsymbol{e}_\phi$ are not necessarily orthogonal.
The surface metric tensor on the nanospring, $g_{ij}= \boldsymbol{e}_i \cdot \boldsymbol{e}_j (i,j = \theta, \phi)$,
and its determinant are calculated as
\begin{gather}
	g_{\theta \theta} = a^2, \, g_{\theta \phi} = g_{\phi \theta} = -\tau a \gamma \sin \theta, \,
	g_{\phi \phi} = \gamma^2 ( 1 + \xi^2), \,
	\nonumber \\
	g \equiv \det g_{ij} = a^2 \gamma^2 \eta^2 .
\end{gather}
Hereafter, we use the dimensionless parameters $\widetilde{R} \equiv R/\gamma$ and $\widetilde{a} \equiv a/\gamma$.

\subsection{Quantities needed for Pauli equation for nanospring}

A curved geometry is known to induce the geometric potential\cite{bib:1649} in general,
which affects the dynamics of a particle confined to the curved surface,
whether the dynamics is relativistic or not.
This potential allows the dynamics to be distinct from that on a flat surface.
The geometric potential for an electron with effective mass $m$ on the nanospring is calculated (see Appendix A for details) as
\begin{gather}
	V_S (\theta) 
		= - \frac{1}{2m a^2} \Bigg[
			\frac{\{ \xi (1 + \xi^2) + \widetilde{a} ( 2 \cos^2 \theta + \xi^2) \sin \theta \}^2}{4 \eta^6}
	\nonumber \\
			- \frac{\widetilde{a} ( \xi^3 \sin \theta - \widetilde{a} \cos^4 \theta )}{\eta^4} 
		\Bigg]
	,
	\label{geopot_spring}
\end{gather}
independent of $\phi$.
$V_S$ takes extreme values on the outer rim ($\theta = \pi/2$) and the inner rim ($\theta = 3 \pi/2$) of the nanospring.

To derive the Pauli equation for the nanospring, we have to calculate the $h$ matrices\cite{bib:1893},
which are responsible for the spin-dependent part of the equation.
They are calculated as
\begin{gather}
	h^{\theta \phi} = 
		\frac{1}{a \gamma \eta^2}
		\begin{pmatrix}
			\xi \cos \theta & i e^{-i \phi} \zeta \\
			-i e^{i \phi} \zeta^* & -\xi \cos \theta \\
		\end{pmatrix}
	, \nonumber \\
	h^{n \theta} =
		\frac{1}{a \eta}
		\begin{pmatrix}
			\tau & -i e^{-i \phi} \xi \\
			i e^{i \phi} \xi & -\tau \\
		\end{pmatrix}
	, \nonumber \\
	h^{\phi n} =
		\frac{1}{\gamma \eta}
		\begin{pmatrix}
			-\sin \theta & e^{-i \phi} \cos \theta  \\
			e^{i \phi} \cos \theta  & \sin \theta \\
		\end{pmatrix}
	,
	\label{hmat_spring}
\end{gather}
where $\zeta (\theta) \equiv \tau \cos \theta - i \xi \sin \theta$.
$h^{ij}$ is antisymmetric with respect to the superscripts.
For details of their derivation, see Appendix B.

\section{Pauli Equation for Nanospring}

As an extension of the Schr\"odinger equation for a curved surface under an electromagnetic field derived by Ferrari and Cuoghi\cite{bib:1645},
the Pauli equation for the curved surface was provided recently by the author.\cite{bib:1893}
We derive, in this section, the Pauli equation for the nanospring as a specific case.

\subsection{Pauli equation in ordinary three-dimensional space}

Here we briefly review the Pauli equation in the ordinary three-dimensional space.
It was originally derived by expanding the Dirac equation for a spin-$1/2$ particle under an electromagnetic field using the Foldy-Wouthuysen method,\cite{bib:152}
which is for the upper two components $\psi$ of the four-component spinor.
In the present study, we use the charge $-e$ and the gyromagnetic factor $\widetilde{g}$ of an electron on the nanospring.
The Pauli Hamiltonian, which neglects the mass term $mc^2$ and the terms on the orders higher than $m^{-2}$, takes the form
\begin{gather}
	H_\mathrm{P} =
	\frac{\Pi^2}{2m} + V
	+ \widetilde{g} \mu_\mathrm{B} \boldsymbol{S} \cdot \boldsymbol{B}
	\nonumber \\
	+ \frac{e}{4m^2c^2} [
		\boldsymbol{\Pi} \cdot \boldsymbol{S} \times \boldsymbol{E}
		+ \boldsymbol{S} \times \boldsymbol{E} \cdot \boldsymbol{\Pi}
	]
	.
	\label{Pauli_Cartesian}
\end{gather}
$\boldsymbol{\Pi} \equiv \boldsymbol{p} + \frac{e}{c} \boldsymbol{A}$ is the kinetic momentum operator and the magnetic field $\boldsymbol{B} = \nabla \times \boldsymbol{A}$ is the rotation of the vector potential.
$\boldsymbol{S} = \boldsymbol{\sigma}/2$ is the spin operator and $\boldsymbol{\sigma}$ is the Pauli matrix.
$V$ is the electrostatic potential.
$\mu_\mathrm{B}$ is the Bohr magneton.
We identify $m$ not with the bare mass $m_e$, but with the effective mass of an electron.
The quantization axis of spin is taken to be parallel to the $z$ axis throughout the present study.

\subsection{Pauli equation and helical momentum of nanospring}

The author derived\cite{bib:1893} the Pauli equation for a curved surface under an electromagnetic field
by starting from the Pauli equation in the ordinary three-dimensional space of the form eq. (\ref{Pauli_Cartesian}).
In the present study, we assume the electrostatic potential $V$ to contain only the confinement part,
which does not explicitly appear in the Pauli equation for the curved surface.
The only potential felt by an electron moving on the nanospring is thus the geometric potential $V_S$, eq. (\ref{geopot_spring}).
The electric field is set to be perpendicular to the nanospring with a uniform strength,
and the magnetic field is set to be static and along the $z$ direction:
$\boldsymbol{E} = E \boldsymbol{e}_n$ and
$\boldsymbol{B} = B \boldsymbol{e}_{(z)}$.
The electric field breaks the inversion symmetry on the surface and will be the origin of the SOI taking place in the present system.
The magnetic field is realized by adopting the vector potential in the symmetric gauge,
$\boldsymbol{A} = (B/2) ( -y \boldsymbol{e}_{(x)} + x \boldsymbol{e}_{(y)} )$,
which has the following components with respect to the curvilinear coordinates,
\begin{gather}
	A_{\theta} = 0, \,
	A_{\phi} = \frac{B \gamma^2 \xi^2}{2}, \,
	A_n = -\tau \frac{B \gamma \xi \sin \beta}{2 \eta}
	,
\end{gather}
on the surface of the nanospring.
Since $A_n$ is nonzero, for establishing the surface Pauli equation, it is necessary to perform a gauge transformation\cite{bib:1645,bib:1893} so that $A_n$ and its normal derivative vanish on the surface.
Such a gauge transformation is always possible and $A_\theta$ and $A_\phi$ are unchanged via the transformation.
The surface wave function $\chi (\theta, \phi)$ satisfies
the time-independent Pauli equation for the curved surface as $H \chi = \varepsilon \chi$,
where $H$ is the surface Pauli Hamiltonian and $\varepsilon$ is the energy eigenvalue.
$H$ is decomposed into the spin-independent and spin-dependent parts as $H = H_{\mathrm{Sch}} + H_{\mathrm{sp-rel}}$,
which are given by\cite{bib:1893} 
\begin{gather}
	H = H_{\mathrm{Sch}} + H_{\mathrm{sp-rel}}
	\\
	H_{\mathrm{Sch}} = -\frac{1}{2m} \Bigg[
		\frac{1 + \xi^2}{a^2 \eta^2} \partial_\theta^2 
		+ \tau \frac{2 \sin \theta}{a \gamma \eta^2} \partial_\theta \partial_\phi 
		+ \frac{1}{\gamma^2 \eta^2} \partial_\phi^2
	\nonumber \\
		+ \frac{\cos \theta}{a^2 \eta^4} \Bigg\{
			\frac{a}{\gamma} \xi (\xi^2 + \cos 2 \theta) + (1 + \xi^2) \sin \theta
		\Bigg\} \partial_\theta 
	\nonumber \\
		+ \tau \frac{\cos \theta}{a \gamma \eta^4}
			\Bigg(
				1 + \xi^2 - \frac{a}{\gamma} \xi \sin \theta
			\Bigg)
			\partial_\phi
	\nonumber \\
		+ \frac{ie}{c} \tau \frac{B \gamma \xi \cos \theta}{2 a \eta^4}
		\Bigg\{
			\xi (1 + \xi^2) + \frac{a}{\gamma} (\xi^2 + 2 \cos^2 \theta) \sin \theta
		\Bigg\}
	\nonumber \\
		+ \frac{2ie}{c} \frac{B \gamma \xi^2}{2 a \eta^2} \Bigg( \tau \sin \theta \partial_\theta + \frac{a}{\gamma} \partial_\phi \Bigg)
		- \frac{e^2}{c^2} \frac{B^2 \gamma^2 \xi^4}{4 \eta^2} 
	\Bigg] + V_S
	,
	\label{Hsch_spring}
	\\
	H_{\textrm{sp-rel}} = 
		\frac{\widetilde{g} \mu_\mathrm{B} B}{2} \sigma_{(z)}
	- \frac{i \alpha_\mathrm{R}}{a \eta} 
	\Bigg[
		\begin{pmatrix}
			\tau    &  -i e^{-i \phi}  \xi \\
			i e^{i \phi}   \xi  & -\tau 
		\end{pmatrix}
		\partial_\theta
	\nonumber \\
		+
		\frac{a}{\gamma}
		\begin{pmatrix}
			 \sin \theta   &  -e^{-i \phi} \cos \theta  \\
			-e^{i \phi} \cos \theta & -  \sin \theta
		\end{pmatrix}
		D_{\phi}
	\Bigg]
	\label{Hsp-rel_spring}
	,
\end{gather}
where $D_{\phi} = \partial_{\phi} + i \frac{eB \gamma^2 \xi^2}{2c}$ is the covariant derivative.
$\alpha_\mathrm{R} \equiv \frac{eE}{4 m^2 c^2}$ is the Rashba coefficient.\cite{bib:Rashba}
Owing to the nontrivial metric of the surface, the normalization condition of the surface wave function should be set as
\begin{gather}
	\int \int \sqrt{g} \diff \theta \diff \phi \, \chi^\dagger \chi = \mathrm{const}.
	\label{normcond_chi}
\end{gather}

\subsection{Helical momentum and its eigenstates}

In the present case, the electron traveling on the nanospring is inevitably forced to move circularly around the axis of the spring.
Thus its movement can convey an orbital angular momentum.
Here we define the helical momentum operator as
\begin{gather}
	K_z \equiv L_z + S_z + \tau \gamma p_z
	\nonumber \\
		= -i(x \partial_y - y \partial_x) + \frac{\sigma_{(z)}}{2} -i \tau \gamma \partial_z
		= -i \partial_\phi + \frac{\sigma_{(z)}}{2}
	,
\end{gather}
where we have used eq. (\ref{spring_point}) for obtaining the last equality.
It is easily confirmed that $K_z$ commutes with the Pauli Hamiltonian $([ H, K_z] = 0)$,
and we can thus obtain simultaneous eigenstates of the energy and the helical momentum.
We therefore put the solution of the Pauli equation in the form
\begin{gather}
	\chi_\nu (\theta, \phi) = \frac{g^{-1/4}}{\sqrt{2 \pi}}
	\begin{pmatrix}
		e^{i (\nu - 1/2) \phi} \psi_{\nu \uparrow} (\theta) \\
		e^{i (\nu + 1/2) \phi} \psi_{\nu \downarrow} (\theta)
	\end{pmatrix}
	,
	\label{chi_nu}
\end{gather}
where $\nu$ is real and can take continuous values.
$\chi_\nu$ is an eigenstate of the helical momentum with the eigenvalue $\nu$: $K_z \chi_\nu = \nu \chi_\nu$.
It is clear from the definition of $K_\nu$ that the electron's circular movement in the positive (negative) direction of $\phi$ provides positive (negative) contribution to $\nu$ regardless of the chirality $\tau$.
We adopt the normalization condition such that one electron resides in the nanospring per twist
and the normalization condition, eq. (\ref{normcond_chi}), becomes
\begin{gather}
	\int_0^{2 \pi} \diff \theta \,  ( | \psi_{\nu \uparrow} |^2 + | \psi_{\nu \downarrow} |^2 ) = 1
	.
	\label{normcond_psi}
\end{gather}
Whereas $\chi_\nu$ has the dimension of inverse length, $\psi_\nu$ is dimensionless.
From eqs. (\ref{Hsch_spring}), (\ref{Hsp-rel_spring}), and (\ref{chi_nu}),
the Pauli equation becomes the following one-variable differential equation for the two-component wave function $\psi_\nu$ with dimensionless parameters:
\begin{gather}
	\widetilde{H}_\nu \psi_\nu = \widetilde{\varepsilon}_\nu \psi_\nu
	.
	\label{Pauli_psi}
\end{gather}
The $\nu$-fixed Hamiltonian is given by
\begin{gather}
	\widetilde{H}_\nu =
		- \frac{1 + \xi^2}{\eta^2} \frac{\diff^2}{\diff \theta^2} +
		\begin{pmatrix}
			\widetilde{V}_{\nu - 1/2}^{(1)} & 0 \\
			0 & \widetilde{V}_{\nu + 1/2}^{(1)} 
		\end{pmatrix}
		\frac{\diff}{\diff \theta}
	\nonumber \\
		+
		\begin{pmatrix}
			\widetilde{V}_{\nu - 1/2}^{(0)}  & 0 \\
			0 &  \widetilde{V}_{\nu + 1/2}^{(0)} 
		\end{pmatrix}
		+ \widetilde{B}_\mathrm{Z} \sigma_{(z)}
		+ \widetilde{\alpha}_\mathrm{R} \Bigg( \widetilde{W}^{(1)} \frac{\diff}{\diff \theta} + \widetilde{W}_\nu^{(0)} \Bigg)
	,
	\label{H_nu}
\end{gather}
where
\begin{gather}
	\widetilde{V}_\nu^{(1)} =
		- \frac{\sin 2 \theta}{\eta^4}
		( 1 + \xi^2 - \widetilde{a} \xi \sin \theta )
		- i \tau  \frac{2 \widetilde{a} ( \nu + \widetilde{B}_\gamma \xi^2)}{\eta^2} \sin \theta  
	,
	\label{V_nu_1} 
	\\
	\widetilde{V}_\nu^{(0)} =
		- \frac{1}{4 \eta^6} [
			2 \widetilde{a} \sin \theta \xi^5
			+ \{ ( \widetilde{a}^2 + 4) \cos^2 \theta - 2 \} \xi^4
	\nonumber \\
			+ 2 \widetilde{a} ( 1 - 4 \cos^2 \theta ) \sin \theta \xi^3
	\nonumber \\
			+ \{ (-6 \widetilde{a}^2 - 3) \cos^4 \theta + ( 5 \widetilde{a}^2 + 9) \cos^2 \theta - 2  \} \xi^2
	\nonumber \\
			+ 4 \widetilde{a} \cos^2 \theta ( \cos^2 \theta - 3) \sin \theta \xi
			- (2 \widetilde{a}^2 + 3) \cos^4 \theta + 5 \cos^2 \theta
		]
	\nonumber \\
		+ i \tau \frac{ \widetilde{a} \cos \theta}{ \eta^4}  
		\Bigg[
			2 \widetilde{a} \xi \sin \theta ( \nu - \widetilde{B}_\gamma \cos^2 \theta) 
	\nonumber \\
			- ( \nu + \widetilde{B}_\gamma \xi^2) ( 1 + \sin^2 \theta + \xi^2 )
		\Bigg] 
	\nonumber \\
		+ \frac{\widetilde{a}^2 ( \nu + \widetilde{B}_\gamma \xi^2)^2}{\eta^2}
		+ 2m a^2 V_S
	,
	\label{V_nu_0} 
	\\
	\widetilde{W}_\nu^{(0)} =
		i \frac{\cos \theta}{2 \eta^3} ( \widetilde{a} \xi - \sin \theta ) 
		\begin{pmatrix}
			\tau &  -i \xi \\
			i \xi &  -\tau 
		\end{pmatrix}
	\nonumber \\
		- \frac{\widetilde{a}}{\eta}
		\begin{pmatrix}
			 - \sin \theta (\nu - \frac{1}{2} + \widetilde{B}_\gamma \xi^2) &  \cos \theta (\nu + \frac{1}{2} + \widetilde{B}_\gamma \xi^2 ) \\
			 \cos \theta (\nu - \frac{1}{2} + \widetilde{B}_\gamma \xi^2) &  \sin \theta (\nu + \frac{1}{2} + \widetilde{B}_\gamma \xi^2 ) 
		\end{pmatrix}
	, 
	\label{W_nu_0} 
	\\
	\widetilde{W}^{(1)} = -i  \frac{1}{\eta}
	\begin{pmatrix}
		\tau &  -i \xi \\
		i \xi &  -\tau 
	\end{pmatrix}
	.
	\label{W_nu_1} 
\end{gather}
We have defined the dimensionless parameters
\begin{gather}
	\widetilde{\alpha}_\mathrm{R} \equiv 2ma \alpha_\mathrm{R}, \,
	\widetilde{B}_\gamma \equiv \frac{eB \gamma^2}{2c}, \,
	\widetilde{B}_\mathrm{Z} \equiv m a^2 \widetilde{g} \mu_\mathrm{B} B
	.
\end{gather}
It is noted that the eigenvalue $\widetilde{\varepsilon}_\nu$ is dimensionless and one can obtain the corresponding energy eigenvalue by multiplying $\widetilde{\varepsilon}_\nu$ by the energy unit $\varepsilon_\mathrm{u} \equiv \frac{1}{2ma^2}$.

From eqs. (\ref{H_nu}) - (\ref{W_nu_1}),
one can confirm that the Hamiltonian for the helical momentum $\nu$ under the magnetic field $B$ and that with the opposite parameters $- \nu$ and $- B$ are related as
$\sigma_{(y)} \widetilde{H}_{-\nu} (-B)^* \sigma_{(y)} = \widetilde{H}_\nu(B)$.
Hence, if $\psi_{\nu}(B)$ is an energy eigenstate belonging to the eigenvalue $\widetilde{\varepsilon}_\nu (B)$,
$\sigma_{(y)} \psi_{\nu}(B)^*$ is an energy eigenstate for $-\nu$ and $-B$ belonging to the same eigenvalue:
\begin{gather}
	\psi_{-\nu}(-B) = \sigma_{(y)} \psi_{\nu}(B)^*,
	\widetilde{\varepsilon}_{-\nu}(-B) = \widetilde{\varepsilon}_{\nu}(B).
	\label{psi_transform_1}
\end{gather}
Since $\sigma_{(y)} \psi_{\nu}(B)^*$ is the time-reversed state of $\psi_{\nu}(B)$, their spin directions are opposite.
The energy dispersion is even with respect to $\nu$ when the magnetic field is absent.

One can also confirm that the Hamiltonians for nanosprings of opposite chiralities are related as
$\widetilde{H}_\nu (-\tau)^*  = \widetilde{H}_\nu(\tau)$,
which implies
\begin{gather}
	\psi_{\nu}(-\tau) = \psi_{\nu}(\tau)^*,
	\widetilde{\varepsilon}_{\nu}(- \tau) = \widetilde{\varepsilon}_{\nu}(\tau).
	\label{psi_transform_2}
\end{gather}

\subsection{Densities and currents of physical quantities on nanospring}

With the definition of the density and the current of an operator for a physical quantity,
the plausible time development equation is obtained for the Pauli equation (see Appendix C).
We define the density functions of the number, spin, and orbital angular momentum of an electron associated with the surface wave function $\chi_\nu$ as
\begin{gather}
	n_\nu  \equiv \chi_\nu^\dagger \chi_\nu
		= \frac{1}{2 \pi a \gamma \eta} ( |\psi_{\nu \uparrow}|^2 + |\psi_{\nu \downarrow}|^2  )
	, \\
	s_{\nu x} \equiv \chi_\nu^\dagger \frac{\sigma_{(x)}}{2} \chi_\nu = \frac{1}{2 \pi a \gamma \eta} \mathrm{Re} (e^{i \phi} \psi_{\nu \uparrow}^* \psi_{\nu \downarrow})
	, \nonumber \\
	s_{\nu y} \equiv \chi_\nu^\dagger \frac{\sigma_{(y)}}{2} \chi_\nu = \frac{1}{2 \pi a \gamma \eta} \mathrm{Im} (e^{i \phi} \psi_{\nu \uparrow}^* \psi_{\nu \downarrow})
	, \nonumber \\
	s_{\nu z} \equiv \chi_\nu^\dagger \frac{\sigma_{(z)}}{2} \chi_\nu = \frac{1}{4 \pi a \gamma \eta} ( |\psi_{\nu \uparrow}|^2 - |\psi_{\nu \downarrow}|^2 )
	,
	\label{def_spdens}
	\\
	l_{\nu z} \equiv \mathrm{Re} [ \chi_\nu^\dagger L_z \chi_\nu ]
	\nonumber \\
	= \frac{\xi^2 }{2 \pi a^2 \eta^3} \Bigg[
		\tau \sin \theta \mathrm{Im} ( \psi_{\nu \uparrow}^* \partial_\theta \psi_{\nu \uparrow} + \psi_{\nu \downarrow}^* \partial_\theta \psi_{\nu \downarrow} )
	\nonumber \\
		+ \frac{a}{\gamma} \Bigg\{
			\Bigg( \nu - \frac{1}{2} \Bigg) |\psi_{\nu \uparrow}|^2
			+ \Bigg( \nu + \frac{1}{2} \Bigg) |\psi_{\nu \downarrow}|^2
		\Bigg\}
	\Bigg]
	,
\end{gather}
respectively.
While $n_\nu$, $s_{\nu z}$, and $l_{\nu z}$ do not depend on $\phi$, $s_{\nu x}$ and $s_{\nu y}$ do.
From the expression of $\boldsymbol{s}_{\nu}$, it is obvious that the spin direction at an arbitrary $\phi$ on the nanospring can be obtained
by rigidly rotating that at $\phi = 0$ with the same $\theta$.
It also means that  $s_{\nu x}$ and $s_{\nu y}$ averaged over $\phi$ vanish.

Considering the definitions of currents, eqs. (\ref{joK_def}) and (\ref{tildejo_def_Pauli}),
we adopt the following definition of the current of an operator $\mathcal{O}$
associated with the surface wave function $\chi_\nu$:
\begin{gather}
	\boldsymbol{j}_{\mathcal{O} \nu} = \boldsymbol{j}_{\mathcal{O} \nu}^\mathrm{K} + \widetilde{\boldsymbol{j}}_{\mathcal{O} \nu}, \\
	\boldsymbol{j}_{\mathcal{O} \nu}^\mathrm{K} \equiv \frac{1}{2m} \mathrm{Re} [ \chi_\nu^\dagger  \boldsymbol{\Pi} \mathcal{O} \chi_\nu + ( \boldsymbol{\Pi} \chi_\nu )^\dagger \mathcal{O} \chi_\nu ], \\
	\widetilde{\boldsymbol{j}}_{\mathcal{O} \nu} \equiv - \frac{e}{4m^2c^2} \boldsymbol{E} \times \mathrm{Re} [ \chi_\nu^\dagger \boldsymbol{\sigma} \mathcal{O} \chi_\nu ].
\end{gather}

Employing the relations between the energy eigenstates, eqs. (\ref{psi_transform_1}) and (\ref{psi_transform_2}),
it is found that the densities and currents of the physical quantities for $(\tau,B,\nu,\phi)$, $(\tau, -B, -\nu, \phi)$, and $(-\tau, B, \nu, -\phi)$ have the same magnitudes.
Their relative signatures are summarized in Table \ref{table_sgn}.

The $\phi$-averaged total electron density is given by
\begin{gather}
	n^\mathrm{tot} (\theta ; \widetilde{\varepsilon}_\mathrm{F}) = \frac{1}{2 \pi} \int_0^{2 \pi} \diff \phi
		\sum_i \int \diff \nu f (\widetilde{\varepsilon}_{\nu i} ;  \widetilde{\varepsilon}_\mathrm{F})  n_{\nu i}
	, 
\end{gather}
where $\widetilde{\varepsilon}_\mathrm{F}$ is the Fermi level as a parameter and $f(\widetilde{\varepsilon} ;  \widetilde{\varepsilon}_\mathrm{F})$ is the Fermi distribution function.
$i$ denotes the branch of the energy spectra.
The $\phi$-averaged total spin density $\boldsymbol{s}^\mathrm{tot}$ and the total orbital angular momentum density $l_z^\mathrm{tot}$ are calculated similarly.
$s_x^\mathrm{tot}$ and $s_y^\mathrm{tot}$ obviously vanish. 
The total number of electrons as a function of the Fermi level is calculated as
\begin{gather}
	N^\mathrm{tot} (\widetilde{\varepsilon}_\mathrm{F}) =
	\int_0^{2 \pi} \diff \theta \sqrt{g}   2 \pi n^\mathrm{tot} (\theta ; \widetilde{\varepsilon}_\mathrm{F})
	.
\end{gather}
The total spin $\boldsymbol{S}^\mathrm{tot}$ and the total orbital angular momentum $L_z^\mathrm{tot}$ are calculated similarly.

We define the net current of the $j$ component of spin at a point as
\begin{gather}
	\boldsymbol{j}_{s_j} (\theta, \phi; \widetilde{\varepsilon}_\mathrm{F}) =
		\sum_i \int \diff \nu f (\widetilde{\varepsilon}_{\nu i} ;  \widetilde{\varepsilon}_\mathrm{F}) 
 		\boldsymbol{j}_{s_{\nu j}}
	.
\end{gather}
The total spin current per pitch is calculated as its integral over $\phi$ and $\theta$ as
\begin{gather}
	\boldsymbol{J}^\mathrm{tot}_{s_j} (\widetilde{\varepsilon}_\mathrm{F}) =
		\int_0^{2 \pi} \sqrt{g} \diff \theta \int_0^{2 \pi} \diff \phi
		\boldsymbol{j}_{s_j} (\theta, \phi; \widetilde{\varepsilon}_\mathrm{F})
	.
\end{gather}

\begin{table}[h]
\begin{center}
\caption{
Relative signatures of densities and currents of physical quantities for $(\tau, -B, -\nu, \phi)$ and $(-\tau, B, \nu, -\phi)$
with respect to those for $(\tau,B,\nu,\phi)$.
$\theta$ and the other parameters are assumed to be fixed.
$\boldsymbol{j}_{\mathrm{c} \nu}$ represents the charge current.
}
\label{table_sgn}
\begin{tabular}{ccc}
\hline\hline
Quantity & $(\tau, -B, -\nu, \phi)$ & $(-\tau, B, \nu, -\phi)$ \\
\hline
$n_\nu$ & $+$ & $+$ \\
$\boldsymbol{s}_\nu$ & $-$ & $x \, +$ \\
 &  & $y \, -$ \\
 &  & $z \, +$ \\
$l_{\nu z}$ & $-$ & $+$ \\
$\boldsymbol{j}_{\mathrm{c} \nu}$ & $-$ & $x \, -$ \\
 &  & $y \, +$ \\
 &  & $z \, -$ \\
$\boldsymbol{j}_{s_{\nu x}}, \boldsymbol{j}_{s_{\nu z}}, \boldsymbol{j}_{l_{\nu z}} $ & $+$ & $x \, -$ \\
 &  & $y \, +$ \\
 &  & $z \, -$ \\
$\boldsymbol{j}_{s_{\nu y}}$ & $+$ & $x \, +$ \\
 &  & $y \, -$ \\
 &  & $z \, +$ \\
\hline\hline
\end{tabular}
\end{center}
\end{table}

\section{Electronic Properties of Nanospring}

In the present study, the Pauli equation for the nanospring, eq. (\ref{Pauli_psi}), is solved numerically.
We adopt the periodic boundary condition with respect to $\theta$,
which is realized by expanding the two-component wave function in plane waves as
\begin{gather}
	\begin{pmatrix}
		\psi_{\nu \uparrow} \\
		\psi_{\nu \downarrow}
	\end{pmatrix}
	= \frac{1}{\sqrt{2 \pi}} \sum_{n = -\infty}^\infty
		\begin{pmatrix}
			c_{n \nu \uparrow} \\
			c_{n \nu \downarrow} 
		\end{pmatrix}
		e^{i n \theta}
	.
\end{gather}
With this expansion, the normalization condition, eq. (\ref{normcond_psi}), becomes
\begin{gather}
	\sum_{n = -\infty}^\infty
		| c_{n \nu \uparrow} |^2 +
		| c_{n \nu \downarrow} |^2
	= 1.
\end{gather}
We take the terms only for $|n| \leq 20$ in the summation, which yields sufficiently converged results.
The energy eigenvalues are thus obtained by diagonalizing the $82$-dimensional ($41$ wave numbers for each of spin-up and -down states) complex matrix for each $\nu$.

\subsection{Energy spectra with no SOI and no magnetic field}

Let us first examine the energy spectra of the nanospring with the SOI and the magnetic field absent ($\alpha_\mathrm{R} = 0$ and $B = 0$).

\begin{figure*}[htbp]
\begin{center}
\includegraphics[keepaspectratio,width=16cm]{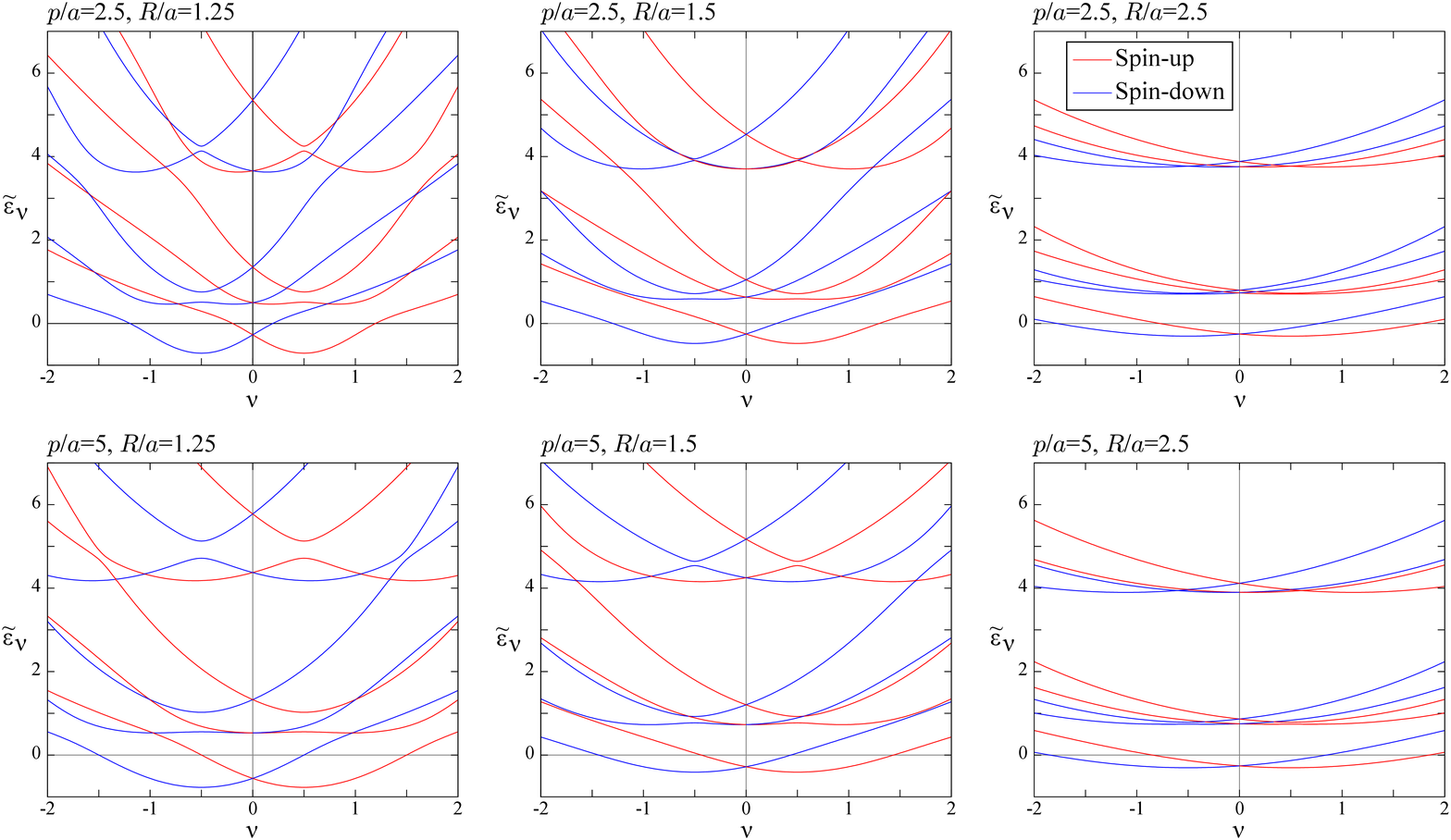}
\end{center}
\caption{
(Color online)
Dimensionless energy spectra of nanospring of $\tau = 1$ for various combinations of $p$ and $R$
as functions of helical momentum $\nu$.
Magnetic field and SOI are absent.
Brighter red and darker blue curves represent purely spin-up and spin-down states, respectively.
}
\label{Fig_energy_nosomag}
\end{figure*}

The dimensionless energies $\widetilde{\varepsilon}_\nu$ of the nanospring of $\tau = 1$ for various combinations of $p$ and $R$
as functions of $\nu$ are plotted in Fig. \ref{Fig_energy_nosomag}.
The variations of the energy spectra due to the change in $R/a$ look larger than those due to the change in $p/a$.
Each of the energy eigenstates is found to be an eigenstate of $\sigma_{(z)}$ unless the energy eigenvalue is degenerate.
Even if the degeneracy occurs, linear combinations of the degenerate energy eigenstates for the eigenstates of $\sigma_{(z)}$ are possible.
This is reasonable since the only contribution that can mix the spin-up and -down components, the SOI, is absent in this case.

In the case of large $R/a$, as shown in Fig. \ref{Fig_energy_nosomag},
the lowest two bands are close to each other and every four out of the higher bands appear in a group.
This observation is understood by considering the limit of large $R$ with $a$ and $p$ fixed,
in which the spatial part of the eigenstate of $\widetilde{H}_\nu$ is of the form $e^{i n \theta}$ with its eigenvalue proportional to $n^2$ apart from a constant, as seen in eq. (\ref{H_nu}).
Hence the lowest group of two bands corresponds to $n = 0$ and each of the other groups of four bands corresponds to $\pm |n| \ne 0$.
The spin degrees of freedom enter in addition and thus such groups of bands are formed.
As seen in Fig. \ref{Fig_energy_nosomag}, the energy dispersion for spin-up (-down) is symmetric around $\nu = 1/2$ $(-1/2)$.
The features of the dispersion will be discussed in detail in the next subsection by considering the transformation laws of the Hamiltonian.

\begin{figure*}[htbp]
\begin{center}
\includegraphics[keepaspectratio,width=16cm]{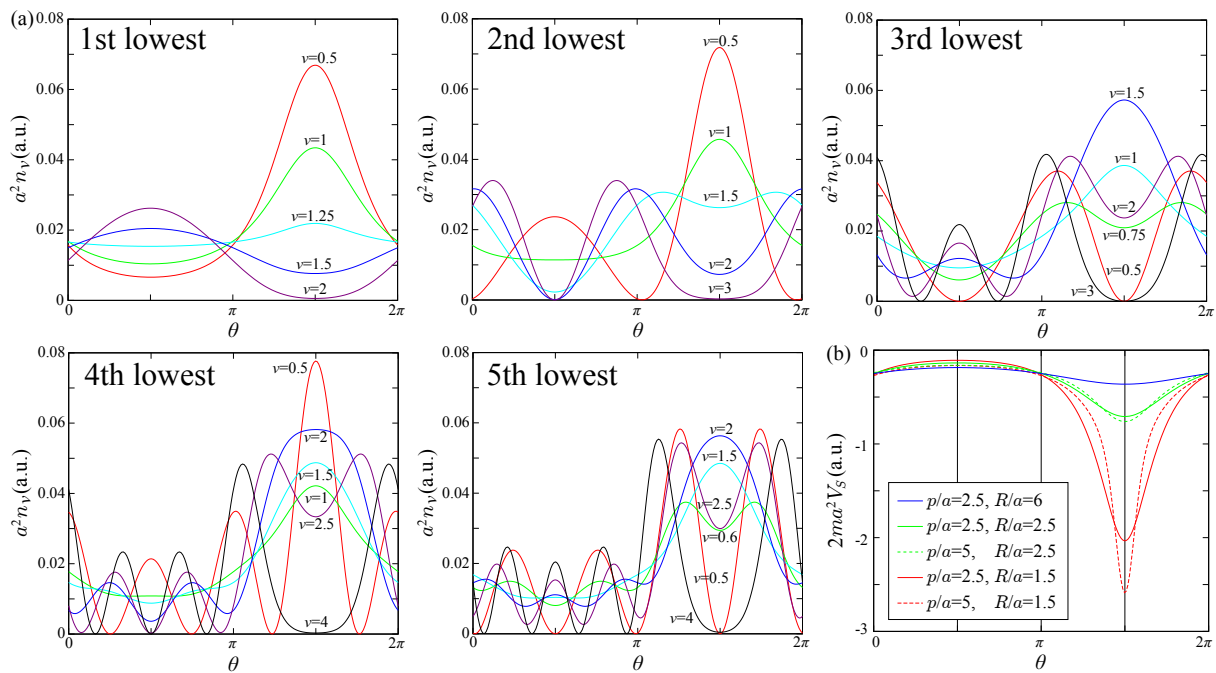}
\end{center}
\caption{
(Color online)
(a) Electron densities $n_\nu$ of the five lowest spin-up energy eigenstates for various values of $\nu$. 
Magnetic field and SOI are absent.
Geometric parameters are fixed at $p/a = 2.5$ and $R/a = 1.5$.
(b) Geometric potential $V_S$ for various combinations of $p$ and $R$.
}
\label{Fig_chg_B0}
\end{figure*}

The electron densities $n_\nu$ of the five lowest spin-up energy eigenstates for various values of $\nu$ are plotted in Fig. \ref{Fig_chg_B0}(a),
with the geometric parameters fixed at $p/a = 2.5$ and $R/a = 1.5$.
Among the eigenstates of a common branch, we observe the tendency that
the electron density  near the axis of the nanospring for large $\nu$ is small compared with that for smalle $\nu$.
This is due to the depletion of the electrons from the inner region caused by centrifugal force.

The geometric potential $V_S$ for various combinations of $p$ and $R$ is plotted in Fig. \ref{Fig_chg_B0}(b).
It is seen that the influence of the change in $R/a$ on the potential is much larger on the inner rim ($\theta = 3 \pi/2$) of the nanospring
than that on the outer rim. ($\theta = \pi/2$) 
The larger $R/a$ is, the weaker the $\theta$ dependence of $V_S$ is.
This is easily understood by considering the asymptotic form of $V_S$.
When $R$ is much larger than $p$ ($\widetilde{R} \gg 1$), the geometric potential becomes
\begin{gather}
	V_S (\theta) \approx 
		\frac{1}{2m a^2} \Bigg[ - \frac{1}{4} + \frac{a}{2 R} \sin \theta \Bigg]
	.
\end{gather}

\subsection{Electronic properties under static magnetic field with no SOI}

Hereafter, we set the geometric parameters of the nanospring as $R = 45$ nm, $a = 30$ nm, and $p = 75$ nm.
In addition, we set the material parameters as $m = 0.05 m_e$ and $\widetilde{g} = -5$,
which are close to the values of bulk InGaAs\cite{bib:1866_14}, for simplicity of our analyses.

\begin{figure}[htbp]
\begin{center}
\includegraphics[keepaspectratio,width=5cm]{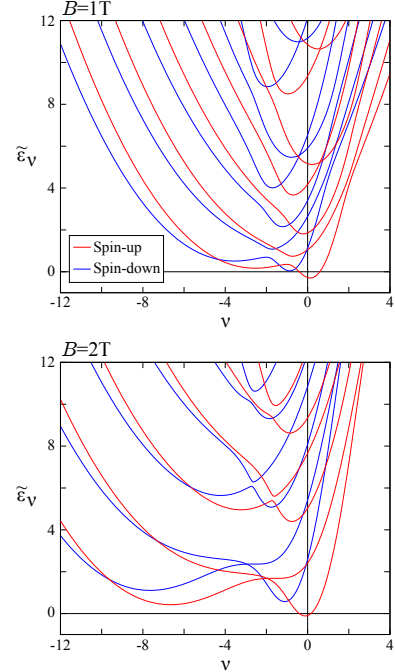}
\end{center}
\caption{
(Color online)
Dimensionless energy spectra of nanospring of $\tau = 1$ for $R = 45$ nm, $a = 30$ nm, and $p = 75$ nm under static magnetic fields.
Material parameters are set as $m = 0.05 m_e$ and $\widetilde{g} = -5$.
Brighter red and darker blue curves represent purely spin-up and spin-down states, respectively.
}
\label{Fig_bandB}
\end{figure}

Let us next examine the electronic properties of the nanospring under a static magnetic field with the SOI absent ($\alpha_\mathrm{R} = 0$ and $B \ne 0$).
The dimensionless energies $\widetilde{\varepsilon}_\nu$ of the nanospring of $\tau = 1$
under magnetic fields $B = 1$ and $2$ T as functions of $\nu$ are plotted in Fig. \ref{Fig_bandB}.
Each of the energy eigenstates is an eigenstate of $\sigma_{(z)}$ as in the cases of $B = 0$ above,
since the SOI is not introduced.
The band dispersion of spin-up (-down) is, however, no longer symmetric around $\nu = 1/2 (-1/2)$, which is due to the nonzero $B$.
The introduction of the magnetic field not only caused the rigid shift of the band dispersion via the Zeeman term,
but also the deformation for the spin-up (-down) states through the terms in $\widetilde{V}^{(1)}_{\nu - 1/2}$ and $\widetilde{V}^{(0)}_{\nu - 1/2}$ ($\widetilde{V}^{(1)}_{\nu + 1/2}$ and $\widetilde{V}^{(0)}_{\nu + 1/2}$) in the Hamiltonian.
It was found that the deformations of the dispersion from the $B = 0$ cases to the $B \ne 0$ cases in Fig. \ref{Fig_bandB}
are caused predominantly by the term $\widetilde{a}^2 ( \nu + \widetilde{B}_\gamma \xi^2)^2/\eta^2$ in $\widetilde{V}^{(0)}_\nu$.
This term acts as an effective potential consisting of three contributions: centrifugal ($\nu^2$), diamagnetic ($B^2$), and cross-term ($\nu B$) ones.
Among them, the cross-term contribution is responsible for the asymmetric shape of the band dispersion of each spin.
As seen in Fig. \ref{Fig_bandB}, the energy eigenvalues of spin-up (-down) states for positive $\nu - 1/2$ ($\nu + 1/2$) are, on the whole, high compared with those for negative $\nu - 1/2$ ($\nu + 1/2$).
This tendency is more obvious for $B = 2$ T than for $B = 1$ T.
This can be understood by considering the classical dynamics of an electron moving at a velocity $\boldsymbol{v}$ on the nanospring under the magnetic field $\boldsymbol{B}$,
due to which the electron feels the Lorentz force $-e \boldsymbol{v} \times \boldsymbol{B}$.
Thus a spin-up electron with positive (negative) $\nu - 1/2$, which corresponds to traveling in the positive (negative) direction of $\phi$ regardless of $\tau$,
feels an attractive force toward (outward) the axis of the spring when $B > 0$,
and acquires a higher (lower) energy due to the centrifugal potential.
This is also the case for a spin-down electron with positive or negative $\nu + 1/2$.
We should keep in mind, however, that this classical interpretation collapses when the spin current on the nanospring is considered, as will be discussed later.

It was confirmed that the asymmetric shapes of the band dispersion in Fig. \ref{Fig_bandB} become their mirror images
when the direction of $\boldsymbol{B}$ is reversed.
On the other hand, their shapes do not change when only the chirality $\tau$ is altered.
These observations are consistent with the relationships of eqs. (\ref{psi_transform_1}) and (\ref{psi_transform_2}).

Let us consider the electronic properties in detail from a mathematical viewpoint.
The Hamiltonian, eq. (\ref{H_nu}), is now diagonal in spin space and the equation to be solved is decoupled for the spin-up and -down components:
\begin{gather}
	\begin{pmatrix}
		\bar{H}_{\nu - 1/2} + \widetilde{B}_\mathrm{Z} & \\
		& \bar{H}_{\nu + 1/2} - \widetilde{B}_\mathrm{Z} \\
	\end{pmatrix}
	\begin{pmatrix}
 		\psi_{\nu \uparrow} \\ 
		\psi_{\nu \downarrow}
	\end{pmatrix}
	= \widetilde{\varepsilon}_\nu
	\begin{pmatrix}
 		\psi_{\nu \uparrow} \\
		\psi_{\nu \downarrow}
	\end{pmatrix}
	,
\end{gather}
where
\begin{gather}
	\bar{H}_{\nu} \equiv
		- \frac{1 + \xi^2}{\eta^2} \frac{\diff^2}{\diff \theta^2} +
			\widetilde{V}_{\nu}^{(1)} \frac{\diff}{\diff \theta}  + \widetilde{V}_{\nu}^{(0)} 
	,
\end{gather}
for which there is the relation
\begin{gather}
	\bar{H}_{-\nu} (-B)^* = \bar{H}_{\nu} (B)
	.
	\label{relation_bH}
\end{gather}
By using a one-component eigenfunction $\bar{\psi}_{\nu \pm 1/2}$ and its corresponding eigenvalue $\bar{\varepsilon}_{\nu \pm 1/2}$ of $\bar{H}_{\nu \pm 1/2}$,
two simultaneous two-component eigenstates of $\widetilde{H}_\nu$ and $\sigma_{(z)}$ can be constructed as
\begin{gather}
	\psi_\nu = 
		\begin{pmatrix}
			\bar{\psi}_{\nu - 1/2} \\
			0
		\end{pmatrix}
		, \,
		\widetilde{\varepsilon}_\nu = \bar{\varepsilon}_{\nu - 1/2} + \widetilde{B}_\mathrm{Z}
	\\
	\psi_\nu = 
		\begin{pmatrix}
			0 \\
			\bar{\psi}_{\nu + 1/2} \\
		\end{pmatrix}
		, \,
		\widetilde{\varepsilon}_\nu = \bar{\varepsilon}_{\nu + 1/2} - \widetilde{B}_\mathrm{Z}
	.
\end{gather}
It is thus clear that the energy dispersion of spin-up states is obtained by shifting that of spin-down states rigidly by $1$ in the $\nu$ direction and $2 \widetilde{B}_\mathrm{Z}$ in the $\widetilde{\varepsilon}_\nu$ direction.
Specifically, if $\psi_\nu$ is a purely spin-up eigenstate for an arbitrary $\nu$ with the eigenvalue $\widetilde{\varepsilon}_\nu$,
the spin-flipped state $\sigma_{(x)} \psi_\nu$ is an eigenstate for $\nu - 1$ with the eigenvalue $\widetilde{\varepsilon}_{\nu - 1} = \widetilde{\varepsilon}_\nu - 2 \widetilde{B}_\mathrm{Z}$.
Similarly,
if $\psi_\nu$ is a purely spin-down eigenstate for an arbitrary $\nu$ with the eigenvalue $\widetilde{\varepsilon}_\nu$,
the spin-flipped state $\sigma_{(x)} \psi_\nu$ is an eigenstate for $\nu + 1$ with the eigenvalue $\widetilde{\varepsilon}_{\nu + 1} = \widetilde{\varepsilon}_\nu + 2 \widetilde{B}_\mathrm{Z}$.

\begin{figure}[htbp]
\begin{center}
\includegraphics[keepaspectratio,width=5cm]{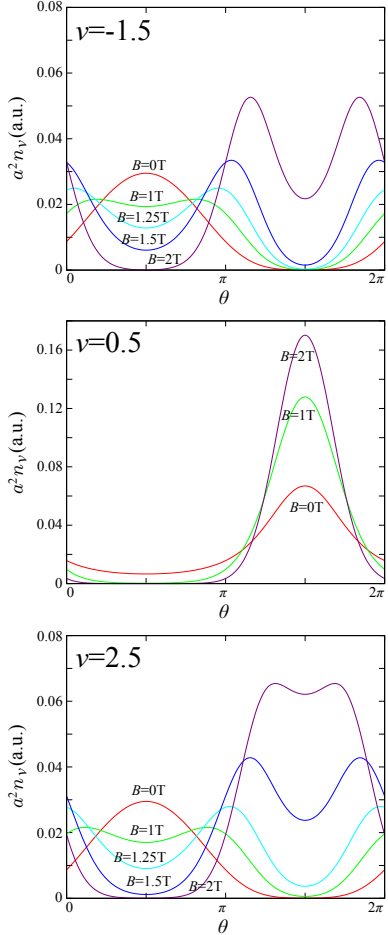}
\end{center}
\caption{
(Color online)
Electron densities $n_\nu$ of the lowest spin-up energy eigenstates for various values of magnetic field $B$. 
SOI is absent.
Geometric parameters are set as $\tau = 1$, $R = 45$ nm, $a = 30$ nm, and $p = 75$ nm.
Material parameters are set as $m = 0.05 m_e$ and $\widetilde{g} = -5$.
}
\label{Fig_chgB}
\end{figure}

The electron densities of the lowest spin-up energy eigenstates for various strengths of the magnetic field are plotted in Fig. \ref{Fig_chgB}.
Although the spin-up electron densities for $\nu = -1.5$ and $2.5$ are the same when $B = 0$ since $\bar{H}_{-2}^* = \bar{H}_2$ [see eq. (\ref{relation_bH})],
when $B \ne 0$, they are different, as clearly seen in Fig. \ref{Fig_chgB}.
The cyclotron movement of the electron with $\nu = 2.5$ is, as discussed above, enhanced by the magnetic field
and thus the localization of the electron density in the inner region of the spring is stronger than that for $\nu = -1.5$.
It is observed that, for all the plotted $\nu$'s, the electron densities spread away the outer region of the spring and pour into the inner region as $B$ increases.
This is because, as the magnetic field becomes stronger, the diamagnetic potential, which is proportional to $B^2$ and pushes the electron toward the inner region,
becomes stronger competing with the centrifugal potential, which pushes the electron toward the outer region.
For the spin-up states of $\nu = 1/2$, the centrifugal potential is absent and thus the localization of the electron density in the inner region of the spring is more significant than those for $\nu \ne 1/2$.

\begin{figure*}[htbp]
\begin{center}
\includegraphics[keepaspectratio,width=16cm]{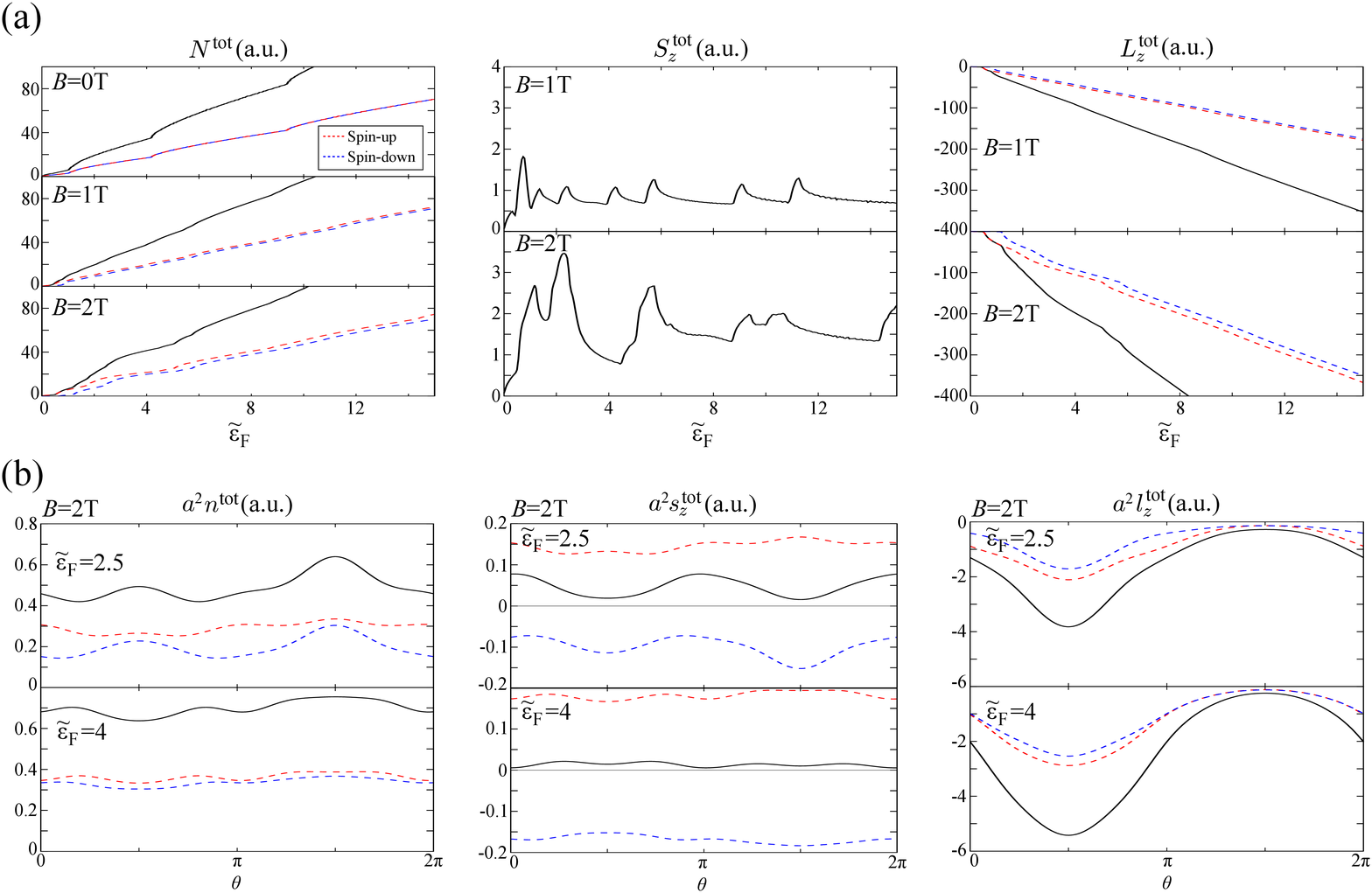}
\end{center}
\caption{
(Color online)
(a) Total number of electrons $N^\mathrm{tot}$, $z$ component of total spin $S_z^\mathrm{tot}$, and $z$ component of total orbital angular momentum $L_z^\mathrm{tot}$
as functions of Fermi level $\widetilde{\varepsilon}_\mathrm{F}$ are plotted as solid curves for various magnetic fields $B$.
Brighter red and darker blue dashed curves represent spin-up and spin-down contributions, respectively.
The origins of energy are set to the lowest eigenvalues for the individual energy spectra.
SOI is absent.
Geometric parameters are set as $\tau = 1$, $R = 45$ nm, $a = 30$ nm, and $p = 75$ nm.
Material parameters are set as $m = 0.05 m_e$ and $\widetilde{g} = -5$.
(b) $\phi$-averaged total number of electrons $n^\mathrm{tot}$, $z$ component of total spin $s_z^\mathrm{tot}$, and $z$ component of total orbital angular momentum $l_z^\mathrm{tot}$
are plotted
as functions of $\theta$ for $\widetilde{\varepsilon}_\mathrm{F} = 2.5$ and $4$ with $B = 2$ T.
}
\label{Fig_densB}
\end{figure*}

The total number of electrons $N^\mathrm{tot}$, the $z$ component of total spin $S_z^\mathrm{tot}$, and the $z$ component of total orbital angular momentum $L_z^\mathrm{tot}$
are plotted in Fig. \ref{Fig_densB}(a)
as functions of the Fermi level $\widetilde{\varepsilon}_\mathrm{F}$.
The eigenstates for spin-up electrons are occupied for $B > 0$ prior to those for spin-down states,
since the spin-up states are energetically more favorable due to the Zeeman effect.
The filling of the bands in such a manner is reflected also in the curves of $S_z^\mathrm{tot}$.
Since the cyclotron movement of the electron in the negative $\phi$ direction is energetically more favorable 
than in the positive one in this case owing to the Lorentz force, as discussed above, the negative $L_z^\mathrm{tot}$ is observed.

The $\phi$-averaged total number of electrons $n^\mathrm{tot}$, the $z$ component of total spin $s_z^\mathrm{tot}$, and the $z$ component of total orbital angular momentum $l_z^\mathrm{tot}$
are plotted in Fig. \ref{Fig_densB}(b)
as functions of $\theta$ for $\widetilde{\varepsilon}_\mathrm{F} = 2.5$ and $4$ with $B = 2$ T.
It is seen that $n^\mathrm{tot}$ is larger around $\theta = 3 \pi/2$ than around $\theta = \pi/2$,
which means that the diamagnetic contribution dominates over the centrifugal contribution in this case.
The contribution to $L_z^\mathrm{tot}$ from the outer region of the spring is much larger than that from the inner region
since the larger distance between the axis and the outer region leads to a larger orbital angular momentum.

\subsection{Electronic properties under static magnetic field with SOI}

\begin{figure*}[htbp]
\begin{center}
\includegraphics[keepaspectratio,width=16cm]{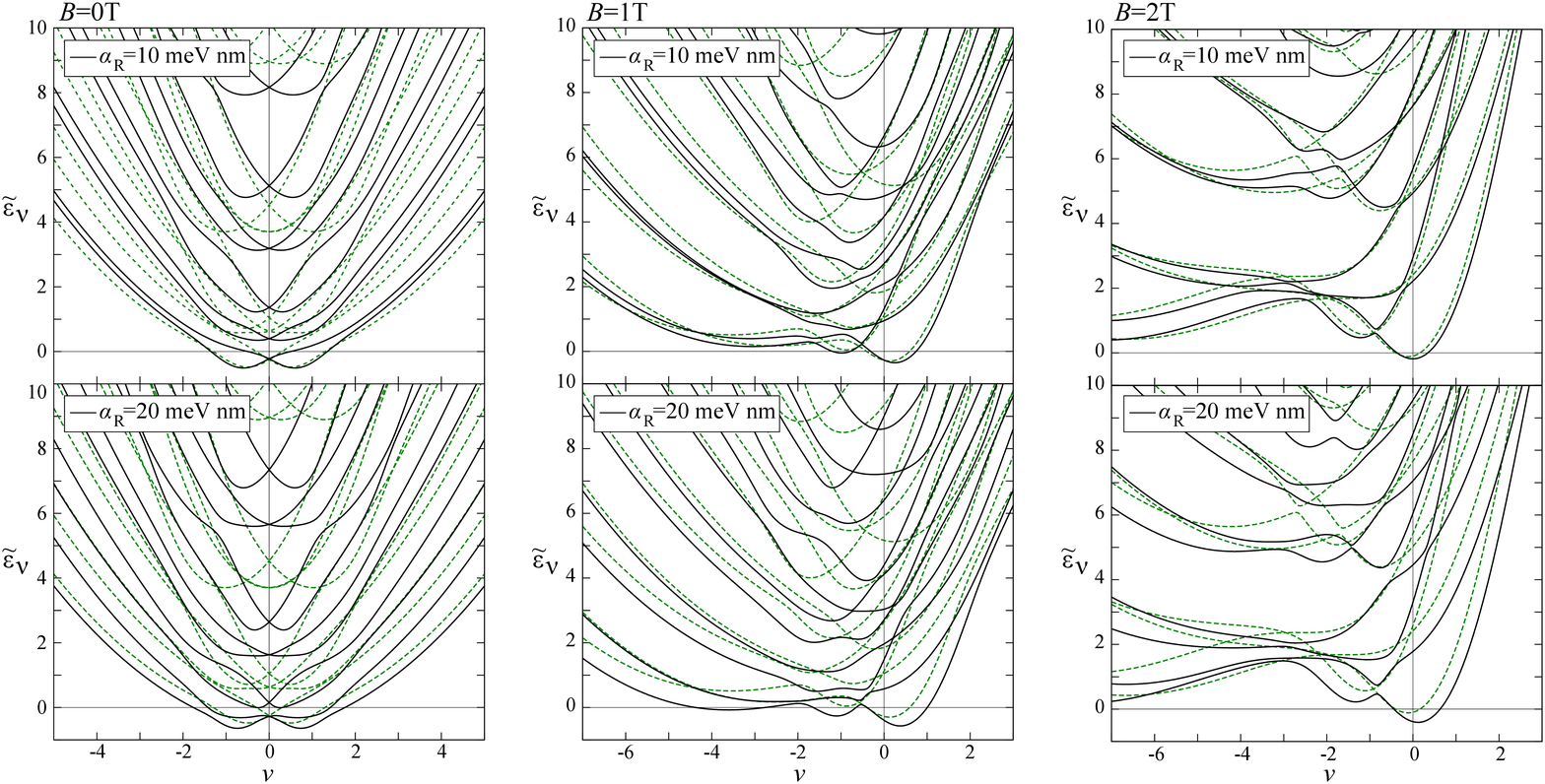}
\end{center}
\caption{
(Color online)
Dimensionless energy spectra for various combinations of $B$ and $\alpha_\mathrm{R}$ as functions of $\nu$.
Dashed curves are for $\alpha_\mathrm{R} = 0$.
Geometric parameters are set as $\tau = 1$, $R = 45$ nm, $a = 30$ nm, and $p = 75$ nm.
Material parameters are set as $m = 0.05 m_e$ and $\widetilde{g} = -5$.
}
\label{Fig_bandBa}
\end{figure*}

Let us now examine the electronic properties of the nanospring under the static magnetic field with the SOI present. ($\alpha_\mathrm{R} \ne 0$ and $B \ne 0$)
The dimensionless energies $\widetilde{\varepsilon}_\nu$ of the nanospring of $\tau = 1$
for various combinations of $B$ and $\alpha_\mathrm{R}$ are plotted in Fig. \ref{Fig_bandBa}.
Some of the degeneracies in the nonrelativistic band structures are resolved by the SOI.
For $\alpha_\mathrm{R} \ne 0$, each of the energy eigenstates is not an eigenstate of $\sigma_{(z)}$
since the SOI mixes the spin-up and spin-down components of the wave function [see eqs. (\ref{W_nu_0}) and (\ref{W_nu_1})].
When the magnetic field is absent, the energy dispersion is even with respect to $\nu$ also for $\alpha_\mathrm{R} \ne 0$ [see eq. (\ref{psi_transform_1})].

\subsection{Persistent spin currents}

From the viewpoint of applications for nanodevices in spintronics,
we are interested particularly in the spin transport along the nanospring.
We therefore examine here the behavior of the persistent spin current occurring on the nanospring in detail
by varying the strengths of the SOI and the external magnetic field.

\begin{figure*}[htbp]
\begin{center}
\includegraphics[keepaspectratio,width=16cm]{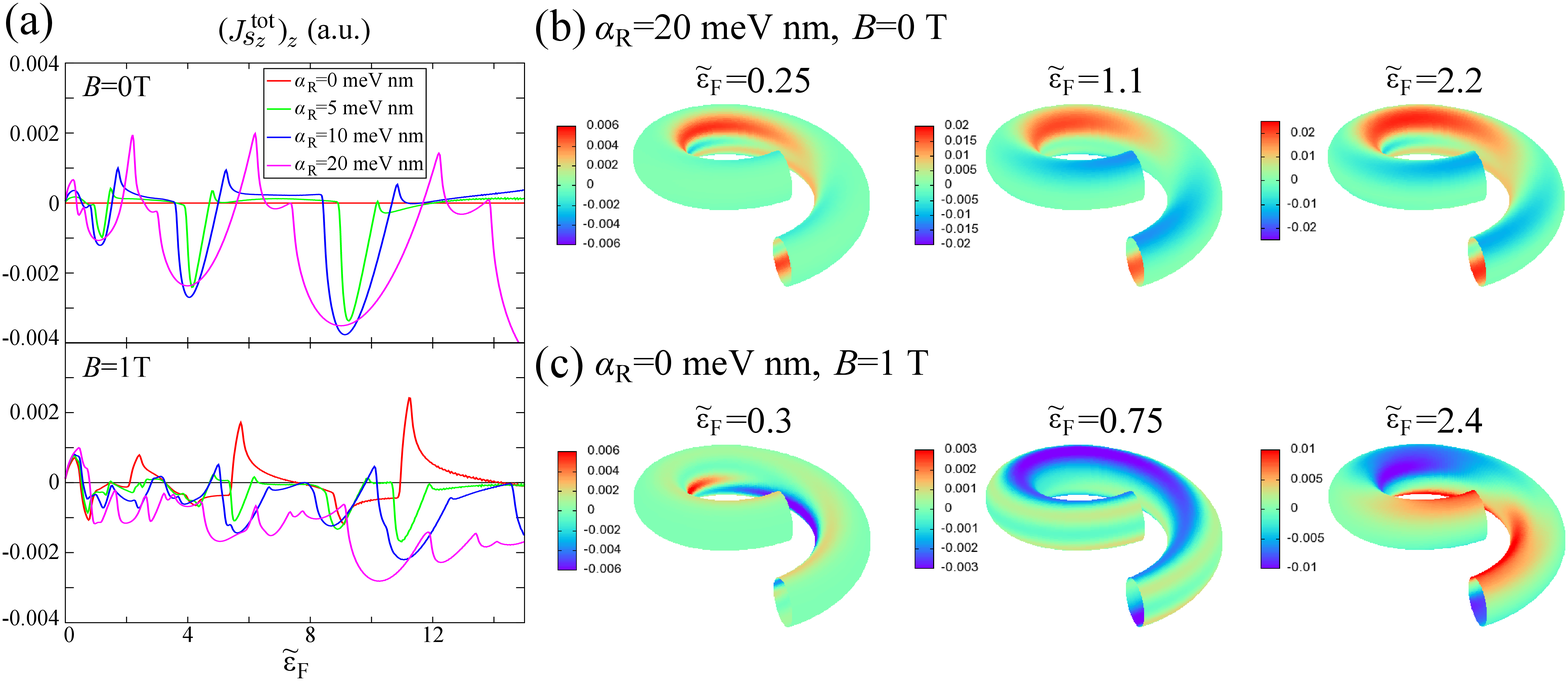}
\end{center}
\caption{
(Color online)
(a) 
$z$ components of $\boldsymbol{J}^{\mathrm{tot}}_{s_z}$ as functions of Fermi level $\widetilde{\varepsilon}_\mathrm{F}$
for various combinations of $\alpha_\mathrm{R}$ and $B$.
The origins of energy are set to the lowest eigenvalues for the individual energy spectra.
Geometric parameters are set as $\tau = 1$, $R = 45$ nm, $a = 30$ nm, and $p = 75$ nm.
Material parameters are set as $m = 0.05 m_e$ and $\widetilde{g} = -5$.
(b) $z$ components of $\boldsymbol{j}_{s_z}$ on the surface of nanospring for $0 \leq \phi < 2 \pi$.
}
\label{Fig_spcurr}
\end{figure*}

The $z$ components of the total spin current $\boldsymbol{J}^{\mathrm{tot}}_{s_z}$ occurring on the nanospring are plotted in Fig. \ref{Fig_spcurr}(a)
for various combinations of $\alpha_{\mathrm{R}}$ and $B$ as functions of the Fermi level.
These results indicate that either a nonzero SOI or a nonzero magnetic field suffices for the occurrence of the persistent spin current.
The $z$ components of $\boldsymbol{J}^{\mathrm{tot}}_{s_x}$ and $\boldsymbol{J}^{\mathrm{tot}}_{s_y}$ were found to vanish irrespective of the values of $\alpha_{\mathrm{R}}$ and $B$.
The $x$ and $y$ components of $\boldsymbol{J}^{\mathrm{tot}}_{s_z}$ also vanish.
$(\boldsymbol{J}^{\mathrm{tot}}_{s_z})_z$ oscillates in a complex manner around the origin as the filling is increased.
This behavior clearly contradicts the interpretation introduced above for the asymmetric features of the band dispersion in Fig. \ref{Fig_energy_nosomag}.
The classical picture employed in the interpretation led to the inequivalence of the positive and negative directions of $\phi$.
If this interpretation were also true for the spin current, the direction of $\boldsymbol{J}^{\mathrm{tot}}_{s_z}$ would not change even when the filling is varied.
To capture the behavior of the spin current in detail, let us observe $(\boldsymbol{j}_{s_z})_z$ on the nanospring, as shown in Fig. \ref{Fig_spcurr}(b).
The regions of positive and negative values exist irrespective of the sign of the total spin current.
This result implies that the oscillatory behavior of the spin current on the nanospring is a quantum mechanical effect that is seen also in other systems.\cite{bib:2061,bib:2069}

The mechanisms for the occurrence of the persistent spin current on the nanospring can be explained qualitatively as follows.

\begin{figure}[htbp]
\begin{center}
\includegraphics[keepaspectratio,width=4cm]{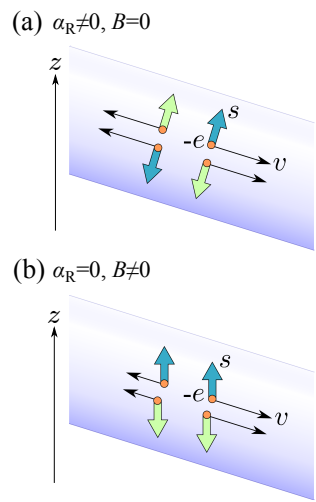}
\end{center}
\caption{
(Color online)
Schematic illustration of the mechanisms for the occurrence of persistent spin current on a nanospring.
(a) When $\alpha_{\mathrm{R}} \ne 0$ and $B = 0$,
two electrons traveling in the same direction with opposite spins are energetically inequivalent.
They give a nonzero contribution to the net spin current.
This is also the case for two electrons traveling in the opposite direction,
whose contribution is the same in magnitude and sign as that from the former two electrons.
(b) When $\alpha_{\mathrm{R}} = 0$ and $B \ne 0$,
two electrons traveling in the same direction with opposite spins are energetically inequivalent.
They give a nonzero contribution to the net spin current.
This is also the case for two electrons traveling in the opposite direction,
whose contribution has the opposite sign to that from the former two electrons
and has a different magnitude.
}
\label{Fig_spcmech}
\end{figure}

We first consider a situation in which the spin-orbit coupling is present and the external magnetic field is absent.
In analogy with the Rashba effect\cite{bib:Rashba} on a flat plane,
two electrons traveling in the same direction with opposite spins are energetically inequivalent due to the SOI, as shown in Fig. \ref{Fig_spcmech}(a).
These two electrons thus give a nonzero contribution to the net spin current.
This is also the case for two electrons traveling in the opposite direction,
whose contribution is the same in magnitude and sign as that from the former two electrons.
The four electrons depicted in Fig. \ref{Fig_spcmech}(a) thus cause a nonzero spin current in total.

We next consider a situation in which the spin-orbit coupling is absent and the external magnetic field is present.
In this case, two electrons traveling in the same direction with opposite spins are energetically inequivalent due to the Zeeman effect, as shown in Fig. \ref{Fig_spcmech}(b).
These two electrons thus give a nonzero contribution to the net spin current.
This is also the case for two electrons traveling in the opposite direction,
and their contribution has the opposite sign to that from the former two electrons
and has a different magnitude due to the inequivalence of the positive and negative $\phi$ directions.
The four electrons depicted in Fig. \ref{Fig_spcmech}(b) thus cause a nonzero spin current in total.

The mechanism in the case of $\alpha_{\mathrm{R}} \ne 0$ and $B = 0$ explained above does not originate in the curvature of the nanospring.
This mechanism is essentially the same as that for the persistent spin current on a flat surface.\cite{bib:2066}
The mechanism for $\alpha_{\mathrm{R}} = 0$ and $B \ne 0$, on the other hand, originates in the curvature of the nanospring,
which couples with the external magnetic field to cause the inequivalence of the orbital motion in opposite directions.

\begin{figure}[htbp]
\begin{center}
\includegraphics[keepaspectratio,width=7cm]{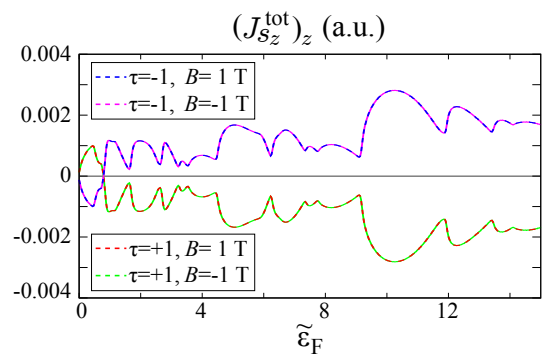}
\end{center}
\caption{
(Color online)
$z$ components of $\boldsymbol{J}^{\mathrm{tot}}_{s_z}$ for four combinations of $\tau$ and $B$ as functions of Fermi level $\widetilde{\varepsilon}_\mathrm{F}$.
Geometric parameters are set as $R = 45$ nm, $a = 30$ nm, and $p = 75$ nm.
Material parameters are set as $m = 0.05 m_e$, $\widetilde{g} = -5$, and $\alpha_\mathrm{R} = 20$ meV nm.
}
\label{Fig_spczz_parity}
\end{figure}

$(\boldsymbol{J}^{\mathrm{tot}}_{s_z})_z$ for four combinations of the chirality and the direction of the external magnetic field are plotted in Fig. \ref{Fig_spczz_parity}.
It is seen that $\boldsymbol{J}^{\mathrm{tot}}_{s_z}$'s for the opposite $\tau$ are in the opposite directions,
while those for the opposite $B$ are in the same direction.
The magnitudes of $\boldsymbol{J}^{\mathrm{tot}}_{s_z}$ were found to be the same for all the four combinations.
This result is a direct consequence of the transformation laws of the wave function [see Table \ref{table_sgn} and eqs. (\ref{psi_transform_1}) and (\ref{psi_transform_2})] originating from the helical geometry of the nanospring.

\section{Conclusions}

In the present study, we first derived the Pauli equation for the nanospring to be satisfied by the two-component wave functions.
The electronic properties of the nanospring are then systematically examined by varying the parameters that characterize the system.
The overall features of the band dispersion were demonstrated to admit the classical interpretation employing the Lorentz force acting on the electron.
The spatial distribution of the electrons on the nanospring was interpreted to be a consequence of the competition between the centrifugal and diamagnetic potentials.
It was demonstrated that either a nonzero SOI or a nonzero external magnetic field suffices for the occurrence of the persistent spin current on the nanospring.
Although we found that the behavior of the spin current on the nanospring does not to allow for the classical interpretation,
we were able to have the simple explanations of the two different mechanisms for the occurrence of the persistent spin current.
One employs the SOI coming from the local inversion asymmetry on the surface,
while the other employs the curvature coupling with the external magnetic field.
A large part of the interesting phenomena observed in the present study comes from the helical geometry,
which specifically means that an electron moving circularly is inevitably forced to travel in the vertical direction.
Similar effects on the electronic properties should thus be observed in other systems in helical or twisted geometries.
The present work will evoke much interest in various curved systems for theoretical and experimental studies in the future.

\begin{acknowledgement}
The present work is fully supported by a Grant-in-Aid for Scientific Research (No. 22104010) from the Ministry of Education, Culture, Sports, Science and
Technology (MEXT), Japan.
\end{acknowledgement}

\appendix

\section{Geometric potential of nanospring}

By taking derivatives of both sides of $\boldsymbol{e}_n \cdot \boldsymbol{e}_n = 1$ with respect to $q_a (q_a = \theta, \phi)$,
we have $\frac{\partial \boldsymbol{e}_n}{\partial q_a} \cdot \boldsymbol{e}_n = 0$.
This means that the derivatives of $\boldsymbol{e}_n$ are linear combinations of the tangent vectors.
Hence, from eq. (\ref{spring_vectors}), we can calculate the Weingarten matrix\cite{bib:1649} $\alpha_{ab}$,
which satisfies $\frac{\partial \boldsymbol{e}_n}{\partial q_a} = \alpha_{ab} \boldsymbol{e}_b$, 
as
\begin{gather}
	\alpha_{\theta \theta} =
		\frac{1}{\eta^3}
			\Bigg[ \frac{\xi (1 + \xi^2)}{a} + \frac{\cos^2 \theta \sin \theta}{\gamma} \Bigg] 
	, \nonumber \\
	\alpha_{\theta \phi} = \tau \frac{1}{\eta^3} \Bigg( \frac{a \cos^2 \theta}{\gamma^2} + \frac{\xi \sin \theta}{\gamma}  \Bigg)
	, \nonumber \\
	\alpha_{\phi \theta} = \tau \frac{1}{\eta a}
	, \,
	\alpha_{\phi \phi} =
		\frac{\sin \theta}{\eta \gamma}
	.
\end{gather}
By substituting $\alpha_{ab}$ into the definition of the geometric potential\cite{bib:1649},
\begin{gather}
	V_S \equiv - \frac{1}{2m} \Bigg[ \frac{(\mathrm{Tr} \, \alpha)^2}{4} - \det \alpha \Bigg]
	,
\end{gather}
that for the nanospring, eq. (\ref{geopot_spring}), is immediately obtained.

\section{$h$ matrices of nanospring}

A point $\boldsymbol{r}$ close to the nanospring is represented by the three coordinates as
\begin{gather}
	\boldsymbol{r}(\theta, \phi, n) = \boldsymbol{r}_{{S}} (\theta, \phi) + n \boldsymbol{e}_n (\theta, \phi)
	,
	\label{r_thpn}
\end{gather}
where $v$ is the distance between the point and the nanospring measured along the normal at $(\theta, \phi)$.
The derivatives of the Cartesian and the curvilinear coordinates with respect to each other,
\begin{gather}
	e_i^{(a)} = \frac{\partial x_{(a)}}{\partial q_i} , \
	e_{(a)}^i = \frac{\partial q_i}{\partial x_{(a)}} 
	,
\end{gather}
satisfy the conditions $e_i^{(a)} e_{(a)}^j = \delta_i^j , e_i^{(a)} e_{(b)}^i = \delta_{(b)}^{(a)}$ due to the chain rule of derivative.
$e_i^{(a)}$ on the nanospring can be calculated from eqs. (\ref{spring_point}) and (\ref{spring_vectors}),
which then enables one to obtain $e_{(a)}^i$ as the inverse of the $3 \times 3$ matrix of $e_i^{(a)}$ as follows:
\begin{gather}
	e_{(x)}^\theta = \frac{(1 + \xi^2) \cos \theta \cos \phi - \tau \xi \sin \theta \sin \phi}{a \eta^2}, \nonumber \\
	e_{(y)}^\theta = \frac{(1 + \xi^2) \cos \theta \sin \phi + \tau \xi \sin \theta \cos \phi}{a \eta^2}, \nonumber \\
	e_{(z)}^\theta = - \frac{\xi^2 \sin \theta}{a \eta^2}, \nonumber \\
	e_{(x)}^\phi = \frac{-\xi \sin \phi + \tau \cos \theta \sin \theta \cos \phi}{\gamma \eta^2}, \nonumber \\
	e_{(y)}^\phi = \frac{ \xi \cos \phi + \tau \cos \theta \sin \theta \sin \phi}{\gamma \eta^2}, \nonumber \\
	e_{(z)}^\phi = \tau \frac{\cos^2 \theta}{\gamma \eta^2}, \nonumber \\
	e_{(x)}^n = \frac{\xi \sin \theta \cos \phi + \tau \cos \theta \sin \phi}{\eta}, \nonumber \\
	e_{(y)}^n = \frac{\xi \sin \theta \sin \phi - \tau \cos \theta \cos \phi}{\eta}, \nonumber \\
	e_{(z)}^n = \frac{\xi \cos \theta}{\eta}
	.
\end{gather}
By substituting these expressions into the definition of the $h$ matrices\cite{bib:1893}, $h^{ij} \equiv \varepsilon_{(abc)} \sigma_{(a)} e_{(b)}^i e_{(c)}^j = -h^{ji}$, where $\varepsilon_{(abc)}$ is the Levi-Civita symbol,
we obtain eq. (\ref{hmat_spring}).

\section{Definition of current of an operator for generic Hamiltonian}

In this Appendix, we look for a possible definition of an arbitrary operator consistent with its time development equation.

Here we consider a Hamiltonian for a two-component spinor $\psi$ of the form
\begin{gather}
	H = \frac{\Pi^2}{2m} + \widetilde{H},
	\label{H_app}
\end{gather}
where $\widetilde{H}$ has the generic form
\begin{gather}
	\widetilde{H} \equiv \sum_{j = 0,x,y,z} \sigma_j [ \boldsymbol{F}_j^{(1)}(\boldsymbol{r}) \cdot \boldsymbol{\Pi} + F_j^{(0)}(\boldsymbol{r}) ]
	.
	\label{Htilde_app}
\end{gather}
$\sigma_0$ is the identity matrix.
$\psi$ obeys the Schr\"odinger equation $i \frac{\partial \psi}{\partial t} = H \psi$.
We assume $\boldsymbol{F}_j^{(1)}$ to be real.
Let us consider the time development of the density function of a physical quantity represented by an operator $\mathcal{O}$,
\begin{gather}
	O(\boldsymbol{r}, t) \equiv \mathrm{Re} [ \psi^\dagger \mathcal{O} \psi ]
	.
\end{gather}
We assume that $\mathcal{O}$ does not depend explicitly on $t$.
The time derivative of this function is, from the Schr\"odinger equation,
\begin{gather}
	i \frac{\partial O}{\partial t} 
		= i \mathrm{Im} \Bigg[
		\psi^\dagger \mathcal{O} \frac{\Pi^2}{2m} \psi - \Bigg( \frac{\Pi^2}{2m} \psi \Bigg)^\dagger \mathcal{O} \psi
	\nonumber \\
		+ \psi^\dagger \mathcal{O} \widetilde{H} \psi - (\widetilde{H} \psi)^\dagger \mathcal{O} \psi
		\Bigg]
	.
	\label{tderiv_odens}
\end{gather}

We define the current associated with $\mathcal{O}$ as
\begin{gather}
	\boldsymbol{j}_\mathcal{O} \equiv \frac{1}{2} \mathrm{Re} [ \psi^\dagger  \boldsymbol{v} \mathcal{O} \psi + ( \boldsymbol{v} \psi )^\dagger \mathcal{O} \psi ]
	\label{def_current_app}
	, 
\end{gather}
where we have defined the velocity operator as $\boldsymbol{v} \equiv -i [ \boldsymbol{r}, H ]$.
From eqs. (\ref{H_app}) and (\ref{Htilde_app}), we have
\begin{gather}
	\boldsymbol{v} = \frac{\boldsymbol{\Pi}}{m} + \sum_{j = 0,x,y,z} \sigma_j \boldsymbol{F}_j^{(1)}
	.
	\label{velo_op}
\end{gather}
Substitution of eq. (\ref{velo_op}) into eq. (\ref{def_current_app}) leads to 
the decomposition of the current into two parts:
$\boldsymbol{j}_\mathcal{O} = \boldsymbol{j}_\mathcal{O}^\mathrm{K} + \widetilde{\boldsymbol{j}}_\mathcal{O}$,
where
\begin{gather}
	\boldsymbol{j}_\mathcal{O}^\mathrm{K} \equiv \frac{1}{2m} \mathrm{Re} [ \psi^\dagger  \boldsymbol{\Pi} \mathcal{O} \psi + ( \boldsymbol{\Pi} \psi )^\dagger \mathcal{O} \psi ],
	\label{joK_def} \\
	\widetilde{\boldsymbol{j}}_\mathcal{O} \equiv  \mathrm{Re} \sum_j \boldsymbol{F}_j^{(1)} \psi^\dagger \sigma_j \mathcal{O} \psi
	.
	\label{tildejo_def}
\end{gather}
The divergence of the kinetic part $\boldsymbol{j}_\mathcal{O}^\mathrm{K}$ is calculated as
\begin{gather}
	\boldsymbol{p} \cdot \boldsymbol{j}_\mathcal{O}^\mathrm{K}
		= \frac{i}{2m} \mathrm{Im} [ \psi^\dagger \Pi^2  \mathcal{O} \psi - ( \Pi^2 \psi)^\dagger \mathcal{O} \psi ]
	\nonumber \\
		= \frac{i}{2m} \mathrm{Im} [ \psi^\dagger \mathcal{O} \Pi^2   \psi - ( \Pi^2 \psi)^\dagger \mathcal{O} \psi ]
		- i \mathrm{Im} \Bigg( \psi^\dagger \Bigg[ \mathcal{O}, \frac{\Pi^2}{2m} \Bigg] \psi \Bigg)
	,
	\label{div_jko}
\end{gather}
and that of the residual part $\widetilde{\boldsymbol{j}}_\mathcal{O}$ is calculated as
\begin{gather}
	\boldsymbol{p} \cdot \widetilde{\boldsymbol{j}}_\mathcal{O}
		= i \mathrm{Im} \Bigg[
			\sum_j - ( \sigma_j \boldsymbol{F}_j^{(1)} \cdot \boldsymbol{p} \psi)^\dagger \mathcal{O} \psi
	\nonumber \\
			+ \psi^\dagger \sigma_j \boldsymbol{F}_j^{(1)} \cdot \boldsymbol{p} \mathcal{O} \psi
			+ ( \boldsymbol{p} \cdot \boldsymbol{F}_j^{(1)} ) \psi^\dagger \sigma_j \mathcal{O} \psi
		\Bigg]
	\nonumber \\
		= i \mathrm{Im} \Bigg[
			\psi^\dagger \mathcal{O} \widetilde{H} \psi - (\widetilde{H} \psi)^\dagger \mathcal{O} \psi
			- \psi^\dagger [ \mathcal{O}, \widetilde{H} ] \psi
	\nonumber \\
			+ \sum_j ( \boldsymbol{p} \cdot \boldsymbol{F}_j^{(1)} - 2i \mathrm{Im} F_j^{(0)}) \psi^\dagger \sigma_j \mathcal{O} \psi
		\Bigg]
	.
	\label{div_tildejo}
\end{gather}
By substituting eqs. (\ref{div_jko}) and (\ref{div_tildejo}) into eq. (\ref{tderiv_odens}),
we obtain
\begin{gather}
	\frac{\partial O}{\partial t}
		= - \nabla \cdot \boldsymbol{j}_\mathcal{O} + \mathrm{Re} [ \psi^\dagger \dot{\mathcal{O}} \psi ]
	\nonumber \\
			+ \sum_{j = 0,x,y,z} (\nabla \cdot \boldsymbol{F}_j^{(1)} + 2 \mathrm{Im} F_j^{(0)}) \mathrm{Re} [ \psi^\dagger \sigma_j \mathcal{O} \psi ]
	,
	\label{current_eq}
\end{gather}
where $\dot{\mathcal{O}} \equiv -i [\mathcal{O}, H]$.
The second and third terms on the right-hand side of this equation are the source terms.
When they are absent, eq. (\ref{current_eq}) is nothing but the continuity equation.

Let us consider here the expressions of currents for the Pauli Hamiltonian, eq. (\ref{Pauli_Cartesian}),
as a special case of the definition, eq. (\ref{def_current_app}).
Comparing eqs. (\ref{Pauli_Cartesian}) and (\ref{Htilde_app}), we find
\begin{gather}
	\boldsymbol{F}_0^{(1)} = \boldsymbol{0}, \,
	F_0^{(0)} = V, 
	\nonumber \\
	{F}_{ij}^{(1)} = -\frac{e}{4m^2c^2} \varepsilon_{ijk} E_k, \,
	F_j^{(0)} = \frac{1}{2} \widetilde{g} \mu_\mathrm{B} B_j
	.
	\label{F_in_Pauli}
\end{gather}
We have used the Maxwell's equation $\nabla \times \boldsymbol{E} = -\frac{1}{c} \frac{\partial \boldsymbol{B}}{\partial t}$,
which vanishes since the magnetic field is static in the present study.
The third term on the right-hand side of eq. (\ref{current_eq}) vanishes in this case.
The residual part of the current hence takes the following form, from eq. (\ref{tildejo_def}),
\begin{gather}
	\widetilde{\boldsymbol{j}}_\mathcal{O} = -\frac{e}{4m^2c^2} \boldsymbol{E} \times \mathrm{Re} [ \psi^\dagger \boldsymbol{\sigma} \mathcal{O} \psi ]
	.
	\label{tildejo_def_Pauli}
\end{gather}
For $\mathcal{O} = 1$, $O = \psi^\dagger \psi$ is the electron density
and its integral over the entire space is the electron number.
It is a conserved quantity since the source term vanishes
and eq. (\ref{current_eq}) reduces to the well-known continuity equation of the probability density.
The integral of the density function of an arbitrary operator is, however, not a conserved quantity in general.
If $\mathcal{O}$ and $H$ commute $(\dot{\mathcal{O}} = 0)$, $\mathcal{O}$ is conserved.

The definition of the spin current is mentioned here.
Since the spin operator is the constant hermitian matrix, the spin current of the form of eq. (\ref{def_current_app}) can be written as
\begin{gather}
	\boldsymbol{j}_{S_j} = \frac{1}{2}  \mathrm{Re} [ \psi^\dagger \{ \boldsymbol{v}, S_j \} \psi ]
	,
\end{gather}
which is nothing but the conventional definition of the spin current.

\end{document}